\documentstyle[11pt,epsf]{article}
\newcommand{\be}{\begin{eqnarray}}
\newcommand{\ee}{\end{eqnarray}}
\newcommand{\tr}{\rm tr}
\newcommand{\tensor}{\otimes}
\newcommand{\dirac}{\not\!\partial}
\title{\begin{flushright}
{\small SUNY-NTG-95-20}
\end{flushright}
{\bf A Mean Field Approach To The Instanton-Induced Effects
Close To The QCD Phase Transition }}
\author{
 {\bf M.Velkovsky and  E.Shuryak  } \\
{\it  Department of Physics,}\\
{\it State University of New York, Stony Brook, NY 11794}}

\begin{document}

\maketitle
\centerline{\bf Abstract}
In the instanton models the chiral phase transition is driven by a transition
from random instanton-antiinstanton liquid and  correlated
instanton-antiinstanton molecules.    So far  this phenomenon was studied  by
numerical simulations, while we develop alternative semi-analytic approach. For
two massless quark flavors, both instantons and ``molecules" generate specific
4-fermion effective interactions. After those are derived, we  determine the
temperature dependence of the thermodynamic quantities, the quark condensate
and the  fraction of molecules using standard mean field method. Using
Bethe-Salpeter equation, we calculate T-dependence of  mesonic  correlation
functions.

\newpage
\section{ Introduction}
  The nature of the phase transition remains one of the main problems
in non-perturbative QCD. Not only it is  important
to understand it from a theoretical point of view, but it is also the
motivation for current and future 
heavy ion collision experiments.
Major efforts are also being made to simulate finite-T QCD on the lattice,
and many results have clarified to some extent the phase diagram of QCD-related 
theories, see reviews  \cite{Karsch,Detar}.

 At zero 
temperature  the mechanism of spontaneous chiral symmetry breaking
 and even the very
existence of most light hadrons is believed  to be connected with the 
fermionic (quasi)zero modes generated by instantons. The specific
instanton liquid model was suggested in \cite{Shu_82}, and then studied 
both analytically
\cite{DP, NVZ} and numerically \cite{Shu_88,SV_90,SSV,SS_95,SS_95b}. 
Even the simplest random model (RILM) was found to reproduce 
mesonic and baryonic correlation functions, 
known from
phenomenology  \cite{Shuryak_cor} and the lattice simulations 
\cite{Negele}. Furthermore, the ``instanton liquid" itself, with parameters
closed to those predicted, were ``distilled" from lattice configurations
by cooling \cite{CGHN_94,MS_95}. 

  However, it is much less
firmly established what happens with instantons at finite temperature T.  
  It has been known for a long time that at $high$ T
the instanton   density is suppressed by the Debye-type screening  
\cite{Shuryak_78,PY},
 and it was thought first that it may be the reason for
chiral symmetry restoration (see e.g. \cite{IS}).
 However both recent theoretical 
analysis at $low$ T  \cite{SMV} and numerical simulations \cite{CS_95}, show 
that up to $T_c$ the instanton density changes are not significant,
 and therefore an instanton
suppression hardly can be an explanation. 

That is why another mechanism was studied in \cite{IS2}:
the phase transition may be driven by ``pairing" of instantons and antiinstantons
into $I\bar I$``molecules". A two-component (or the so called ``cocktail") model
was proposed, in which at $T<T_c$
individual instantons and ``molecules" coexist in ensemble, while (for m=0) at
 $T>T_c$ only molecules survive and the 
chiral symmetry gets restored.

  In \cite{SS_95b} the Cocktail Model was studied numerically, by keeping the 
temperature fixed and changing the molecule fraction
$f$. Many mesonic and baryonic correlation 
functions were calculated: they show very dramatic
 changes as $f$ goes from small values to 1. Furthermore, 
a very interesting hints for survival of some
hadrons $above$ the phase transition were presented.
 This scenario of the phase transition was been recently confirmed by 
numerical simulations of the instanton vacuum at finite temperature 
in which both boson and fermion induced interaction between the 
pseudo-particles is taken into account \cite{SSV_mix}. It was shown
that the  molecules start to form only  close to 
 $T_c$,  and that they are polarized in 
color space as well as in the Euclidean time direction, as anticipated 
in \cite{SS_95,IS2}.

 In the present paper, we investigate the same two-component
(or  ``cocktail") model by different
 (analytic) methods.
The main difference with numerical simulations mentioned above is that
they first calculate quark propagation in a $given$ background field
(a superposition of instantons), averaging over gauge fields $later$.
In our approach we start with the integration over collective 
variables of instantons, deriving {\it effective interaction} between quarks.
Basically, our next steps are the same  as in the
Bardeen-Cooper-Schreiffer theory of superconductivity or Numbu-Jona-Lasinio
model of chiral symmetry breaking. Our 4-fermion interaction
effective 
interaction is, however, much more complicated, 
it includes  two types of nonlocal terms - one from the random 
instantons and antiinstantons, and the another one from the strongly
correlated pairs, or molecules.   

 The paper is organized as follows:

 In section 2 we calculate the effective interaction $S_{eff}$ and the overlap
matrix element $T_{I \bar I}$, at temperature $T$, of an instanton and an 
anti-instanton in the most attractive relative orientation whose separation 
has only a time component (i.e. they lie on the Matsubara circle). After that
we perform an integration over the 11 quasi-zero modes of the molecule to find
the corresponding activity (or internal
partition function of a molecule) $Z_{mol}$. We find out that the equilibrium
configuration of the $I \bar I$ system, in the temperature range around $T_c$, 
is when the instanton and the anti-instanton
  are separated by the maximal  
interval in Matsubara time $\tau={1\over 2T}$,
they have zero spatial separation and are in the most attractive
orientation in color group.

 In section 3, we evaluate the partition function, assuming that
the fermion determinant is a product of the separate instanton, 
antiinstanton and molecule determinants. Next we exponentiate them, by 
performing an Inverse Laplace Transformation. Then we integrate over the 
instanton, antiinstanton and molecule positions to obtain an effective 
four-fermion lagrangian (we study the case of two flavors in the chiral 
limit).

 In section 4, we use this Lagrangian
in a Hartree-Fock approximation, leading to a 
theory of free fermions, but with temperature and  momentum dependent 
mass. Then we calculate the free energy and minimize it in order  
to determine our parameters such as the molecule 
fraction $f$, the quark condensate and the Laplace parameters for the liquid 
and for the molecules. We  solve four mean field equations 
and determine the temperature dependence of all parameters.
The results clearly show the transition from 
instanton liquid to gas of molecules and the restoration of the chiral 
symmetry.

 In section 5, we calculate mesonic correlation functions, 
using the Bethe- Salpeter equation. We also extract the pion-quark couplings, 
the pion decay constant $f_\pi$ and its coupling to pseudoscalar current 
$\lambda_\pi$. All  the results naturally respect the Goldstone theorem 
(for example $f_\pi$ vanishes at $T=T_c$). To 
compare with the numerical simulations for the mesonic 
correlation functions, we perform a Fourier transform to Euclidean time.
Reasonable agreement between them is observed.

\section{ The $I \bar I$ Molecules Around $T_c$}

 This section is relatively independent of the rest of the paper
in the sense that it deals with properties of the
 $single$  $I \bar I$  molecule. It consists of two steps: the first is
 a new calculation of  the $\bar I I$ interaction in the relevant
 configuration
(so to say, the quantum mechanics of it); while the second deals with its
internal statistical
mechanics. In the next sections we will proceed to
the statistical ensemble  of $multiple$ instantons
around the critical temperature.

  The $I \bar I$ interaction and the overlap matrix elements of
the fermionic zero modes, both at zero and at finite temperature, were studied 
in detail in \cite{SV-FT1-91,SSV_95}. Unfortunately these studies 
were not complete in the sense that they did not include our
region of interest -- the configurations in which
 the pseudo-particles lie around
the opposite ends of a diameter on the Matsubara circle. 

The best definition of the $classical$ $ I \bar I$ interaction 
$S_{int}=(S-2S_0),$ where $S_0$ is the single instanton action, is given by
the so-called streamline configurations 
\cite{Ver_91}. Unfortunately, at finite temperature, the
conformal symmetry that allows to find the streamline is
missing. Therefore,
one has to use  some ansatz for the $I \bar I$ configuration which has a
natural extension to non-zero T.
We are going to use the ratio ansatz \cite{Shu_88}
\be
A^a_\mu(x)=-{1 \over g}{O^{ab}_I \bar \eta^b_{\mu \nu}\partial_\nu \Pi_I+
O^{ab}_{\bar I} \eta^b_{\mu \nu}\partial_\nu \Pi_{\bar I} \over \Pi_I+
\Pi_{\bar I}-1},
\ee
where $\Pi_I=\Pi(x-z_I),$ $\Pi_{\bar I}=\Pi(x-z_{\bar I})$ and
\be
\Pi({\bf r},t) = 1+{\pi T \rho^2 \over r}{\sinh (2\pi r T)
\over\cosh (2\pi r T) - \cos( 2\pi t T)},
\ee
 which is free from the artifacts of the sum ansatz and provides a
reasonable repulsive core. The classical bosonic interaction is just
obtained from this expression, by numerical calculation of the
classical action.

The quark-induced interaction is described in general by
 fermionic zero-mode overlap matrix element:
\be
T_{I\bar I}=\int dt d^3{\bf x} \phi_I^\dagger(x-z_I)\not\!\!D\phi_{\bar 
I}(x-z_{\bar I}),
\ee
In this case we use the simplest
sum ansatz \cite{DP_84}, because in this case one can use an
ordinary derivative instead of a covariant one. 
As it was shown in   \cite{SV-FT1-91}, the
results in this case differ insignificantly from the ones obtained by the ratio
ansatz. 
 
The $O(3)$ symmetry of the $I \bar I$ configuration along the Matsubara
circle allows us to reduce the above integral and the integral for the bosonic
part of the action to a two dimensional one, and we perform the integration
numerically. Our results coincide with the ones from \cite{SV-FT1-91,SSV_95}
in the region where their formulas are valid (for Matsubara time separations
that are not close to $L/2={1 \over 2T}$). The results for $T_{I \bar I}$ are
shown in Fig. 1 (a), and the combined $e^{-S_{int}}T^4_{I \bar I}$ in Fig. 1
(b) (the dotted line).

  The second step is to include  the ``internal entropy", or multiplicity
  of different $I \bar I$ configurations into account.
One has to take the integral
\be
Z_{mol}(T)=\int d^{11} \Omega \ T^{2N_f}_{I \bar I}e^{-S_{int}}
\ee
 over all collective variables. These are
the time separation $\tau$, the three dimensional spatial separation
and, for $SU(3)$,
 seven relative orientation angles\footnote{ Note that 
rotations in the direction of $\lambda_8$ does not change the $I \bar I$ 
configuration, so that we have only 7 orientation parameters. The measure
of integration is described in \cite{CB}.}. 
Most of the 11-dimensional integral, however, can be done by a saddle-point
approximation. Both  
$S_{int}$ and $T_{I\bar I}$ decay exponentially in the spatial
separation directions.
The second derivatives of $S_{int}$ over  orientation
angles are big, because they are multiplied by $8 \pi^2/g^2=S_0\sim
10>>1$.
In contrast to that, the angular dependence of $-2N_f\ln(T_{I
\bar I})$ is rather weak, so in  angular integrals we 
treat $T_{I \bar I}$ as a constant. Finally, we find
that the saddle point approximation is not good at all for the integral along
the time separation $\tau$
\footnote{
E.g. the second derivative of $S_{int}-2N_f \ln(T_{I \bar
I}$ along this direction for a time separation of $L/2$, is negative for 
temperatures $T< 230 MeV$, although explicit calculations show that this
configuration is a maximum for $z$ and not a minimum.}. That is why
we perform this last integration explicitly.

  In  Fig. 1 (b), we show the time dependence of
the integrand $Z_{mol}'(\tau)=\int 
d^{10}\Omega \  
T^4_{I \bar I}e^{-S_{int}}$, (the integration over all 
variables but the Matsubara time), at $T_c$, compared to the exponent of the 
classical  and quark- induced interaction $\exp-\{S_{int}-2N_f 
\ln(T_{I \bar I})\}$, which has a maximum at about $\tau=L/4$ and a minimum at 
$\tau=L/2$.
One can see that when the multiplicity of phase space is included, the 
$I \bar I$
molecules are distributed around $\tau=1/2T$. The corresponding time
dependence of  $T_{I \bar I}$ at $T_c$, which is is shown in Fig. 1(a), is 
rather weak around the maximum of $Z'$. This allows us to use its central value
for all molecules that participate in our ``cocktail model". 
After the last integral is done, one  gets
the absolute value of the statistical sum  for  molecules $Z_{mol}$.  
The temperature dependence of $Z_{mol}$ and  
$T_{I\bar I}$ is  shown in Figs. 1 (c), (d).

  How accurate is our saddle point
integration? In the next section we will find that in order
to set the phase transition at
 $T_c= 150 MeV$ one needs  $Z_{mol}$ about 1.8 larger: 
 it means only a $5\%-10\%$ error per integration over each of 
the $I \bar I$ collective parameters. However, we think real
uncertainty 
in $Z_{mol}$
is larger, and it is about one order of 
magnitude (as one can judge from 
including $-4\ln(T_{I\bar I})$ in the second derivative over the
orientation parameters, etc). 
 Nevertheless we think its {\it temperature dependence} is evaluated
reliably,
and we use it below in our calculation of
the thermodynamical properties of the instanton ensemble , the
correlation functions and the pion coupling constants.

\section{ Effective 4-Fermion Interaction }

 Let us start with a list of assumptions and simplifications,
 defining the model 
and the particular parametrization used. We emphasize, that most of them are 
possible only because we are primarily interested in a comparatively narrow 
interval of temperature T around $T_c$, in which significant structural 
changes (the phase transition) takes place.

\begin{itemize}
\item We fix the total pseudo-particle density to be a constant 
 for T up to $T_c=150MeV$, with only composition $f$ changing. We take
$n_t=1 fm^{-4}$ specifically, lacking more accurate number. (None of the 
results change qualitatively if, say, it is modified by a factor of 2.) 

\item Lattice data in \cite{CS_95} indicate, that the 
instanton radius doesn't change much with $T$
till well above the phase transition. So we also fix it at
$\rho=0.34 fm$. (However we are going to express all 
dimensional quantities in units of $\rho$ and its powers anyway.) 

\item In the previous section we have shown that the dominant 
configuration for the $I\bar I$ molecules is the one, in which 
the distance between their centers in the Euclidean time direction is equal to
``half-box" $R=L/2 \equiv 1/2T$, the spatial distance is 0, and they have the
most attractive relative orientation in color space. Although the molecule
distribution has some span around $L/2$ in the time  direction, $T_{I \bar I}$
changes little, which allows us to use one  value of it for all molecules. 

\item  $T_c$ and $S_0$ are related by the requirement 
that all unpaired instantons disappear at $T_c$. This condition can be 
satisfied only in a window around $T_c=150 Mev$. We use the following values: 
$T_c=150 MeV, S_0=9.5$. However there is a big uncertainty in this relation, 
due to the uncertainty of the calculated value of the molecule activity $z$.

\item We are going to ignore all bosonic and fermionic 
interactions between  random instantons and among $I\bar I$ molecules.

\end{itemize}

   Now we proceed to 
calculation of  the effective quark interaction. For one instanton it is 
the well known 't Hooft effective interaction. It can be derived from the 
zero-mode part of the quark propagator :
\be
S(x,y)=S_{NZM}(x,y)+{\phi_I(x-z_I)\phi^\dagger_I(y-z_I)\over -im},
\ee  
where $\phi(x)$ is the quark zero mode in the field of an instanton with 
center at $z_I$ and $S_{NZM}(x,y)$ represent contributions of non-zero modes,
(following \cite{DP} we shall approximate these contributions by the free 
quark propagator $S_0$).
For $N_f$ light fermions, one should take the corresponding  power of $S(x,y)$,
multiply it with the instanton probability (P$\sim m^{N_f}$) and 
then cancel m in both, resulting in the interaction which is finite in the 
chiral limit (m=0).

  Similar procedure can be repeated for an isolated
$I\bar I$ molecule \cite{SSV_mix}. In the chiral limit quarks have the 
following propagator:
\be
 S_m(x,y)=S_0(x,y)+{\phi_{I(m)}(x-(z_m-L/4))
\phi^\dagger_{\bar I(m)}
(y-(z_m+L/4))\over T_{I\bar I}}+ \cr
{\phi_{\bar I(m)}(x-(z_m+L/4))\phi^\dagger_{I(m)}
(y-(z_m-L/4)) \over T_{I\bar I}},
\ee  
where $T_{I\bar I}$ was defined in (2).

 For $N_f=2,$ the following fermionic path integral generates the 
above propagators:
\be
Z=\int d \psi d \psi ^\dagger {\exp \{\int d^4 x \psi ^\dagger i \dirac
\psi\} \over 
N_+!N_-!N_m!}\prod_{I=1}^{N_+}c_{\rho}\theta_+
 \prod_{\bar I=1}^{N_-} c_{\rho}\theta_-  \prod_{m=1}^{N_m}
c_{\rho}^2 c \theta_m,
\ee
where 
\be
\theta_+=\int d z_I d\Omega_I \prod_{f=1}^{2}\big(\int d^4x\psi^
\dagger_f(x) i \dirac \phi_I (x-z_I)
\int d^4y\phi_I^\dagger(y-z_I)i \dirac \psi_f (y)\big), \\ 
\theta_-=\int d z_{\bar I} d\Omega_{\bar I}\prod_{f=1}^{2}\big(
\int d^4x\psi^\dagger_f(x) i \dirac \phi_{\bar I}(x-z_{\bar I})  
\int d^4y\phi_{\bar I}^\dagger(y-z_{\bar I})i \dirac \psi_f(y)\big),\\
\theta_m=\int d z_m d\Omega_m \Big[ \prod_{f=1}^{2} \Big(T_{I \bar I}+
\int d^4x\psi^\dagger_f(x) i \dirac \phi_{I(m)}(x-z_m+{L\over4}) 
\times \qquad \qquad \quad \ \cr
\int d^4y\phi_{\bar I(m)}^\dagger(y- z_m-{L\over4})i \dirac \psi_f
(y)+\int d^4x\psi^\dagger_f(x)i\dirac \phi_{\bar I(m)}(x-z_m-{L\over4})
\times \cr \int d^4y\phi_{I(m)}^\dagger(y-z_m+{L\over4})i \dirac \psi_f 
(y)\Big)+\sum_{f=1}^{2} \Big(\int d^4x\psi^\dagger_f(x) i \dirac 
\phi_{I(m)}(x-z_m+{L\over4}) \times \cr 
\int d^4y\phi_{I(m)}^\dagger(y- z_m+{L\over4})i \dirac \psi_f
(y) \int d^4x'\psi^\dagger_f(x')i\dirac \phi_{\bar I(m)}(x'-z_m-{L\over4})
\times \cr \int d^4y'\phi_{\bar I(m)}^\dagger(y'-z_m-{L\over4})i \dirac \psi_f
(y')\Big)\Big],
\ee
Note, that there are no odd terms in $T_{I \bar I}$, because of the integration
over $\Omega_m$. Here $z_i$ and $\Omega_k$ are the 
collective coordinates of an $I\bar I$ molecule, 
$ c=Z_{mol}/T_{I \bar I}^2$ and $c_{\rho}$ is the 
single instanton partition function \cite{'t H} integrated over the 
instanton radius $\rho$. The convergence of the latter is due to the 
non-perturbative $g(\rho)$, which tends to a constant for large  $\rho$ 
\cite{MS_95,Shuryak_com}.

Now, applying inverse Laplace Transformation, we get the following  partition
function: 
\be
 Z=const \int d\beta_+d\beta_-d\beta_m 
\exp\{-(N_++1)\log(\beta_+/c_{\rho})-  (N_-+1)\log(\beta_-/c_{\rho}) \cr
-(N_m+1)\log (\beta_m /(c_{\rho}^2 c)) +\int d^4 x (\psi ^\dagger i \dirac
\psi\ -  \beta_+\theta_++\beta_-\theta_-+\beta_m\theta_m )\}. 
\ee 
 To evaluate
the effective interaction $\theta_i$ terms we go to momentum  space and do the 
integrations over the orientation angles and the center coordinates of  the
instantons, antiinstantons and molecules\footnote{ Let us again remind that 
for the molecules we assume complete polarization both in coordinate  and color
space.}. The Fourier transforms of the fermion zero  modes at finite
temperature are \cite{NVZ-ft}: 
\be 
\alpha({\bf k}, \omega_n)=({\bf k}^2+
\omega_n^2)(A^2+B^2),  \ee \be A&=&{1 \over 2\pi \rho}\int_0^{\infty} 4 \pi
r^2dr\int_0^{1/T} d t  \left({\cos( kr) \over kr}-{\sin (kr) \over
k^2r^2}\right)\cos(\omega_n  t)\Pi^{1/2} \times \cr && \hbox{\hskip 2.5 in}
\partial_r \left({(\Pi-1)  \over \Pi}{\cos \pi t T \over \cosh \pi r T}\right)
, \\  B&=&{1 \over 2\pi \rho}\int_0^{\infty} 4 \pi r^2dr\int_0^{1/T} d t 
{\sin( kr) \over kr}\sin(\omega_n t)\Pi^{1/2}\partial_r \left({(\Pi-1) \over 
\Pi}{\cos \pi t T \over \cosh \pi r T}\right) , 
\ee 
where $\Pi$ is the
potential (2) that also appears in the  finite  temperature zero modes 
$\phi^R$ in coordinate space \cite{GPY}: 
\be 
\phi^R={1\over 2
\pi\rho}\Pi^{1/2}\dirac \left({(\Pi-1) \over  \Pi}{\cos \pi t T \over \cosh \pi
r T}\right) {\bf \epsilon}^R, 
\ee
and ${\bf \epsilon}^R$ is the standard
constant spinor coupling spinor  and Dirac indexes. Combining all fermion  
terms together, one obtains the following  nonlocal four-fermion interaction
\footnote{All
expressions  throughout the text should be understood as given at finite 
temperature even if we didn't write them explicitly in this form for 
simplicity of the notations, e.g. $\int d^4 k/(2 \pi)^4 $ means$T\sum 
_{n=-\infty}^{\infty}\int d^3 {\bf k}/(2 \pi)^3$, etc. }:  

\be 
S_{4f}= \int\prod_{i=1}^4{d^4 k_i \over 
(2\pi)^4}(2\pi)^4\delta^4(k_1-k_2+k_3-k_4)\sqrt{\alpha(k_1)\alpha(k_2)}  \times
\cr \sqrt{\alpha(k_3)\alpha(k_4)} \Big\{(\beta_++\beta_-){1\over 
16N_c^2}\left[(\psi^\dagger \tau^-_a \psi)^2-(\psi^\dagger \tau^-_a  i
\gamma_5\psi)^2+ octet\ terms \right]-  \cr \beta_m \bigg[{1\over 4 N_c^2}
\left((\psi^\dagger \tau_a \psi)^2+(\psi^\dagger  \tau_a i \gamma_5\psi)^2
\right) -{1 \over 4 N_c^2}\left((\psi^\dagger  \tau_a \gamma_0
\psi)^2+(\psi^\dagger \tau_a\gamma_0  \gamma_5\psi)^2 \right)+ \cr {1 \over
N_c^2}(\psi^\dagger \gamma_0  \gamma_5\psi)^2+ octet\ terms \bigg] \Big\}, 
\ee
where $\tau^-_a=(i,{\bf  \tau})$, while $\tau_a=(1,{\bf \tau})$.
The four fermion interaction is given in a Fierz symmetric  form, which
means that all permutations of these terms are included. We have  shown 
explicitly only the color singlet terms.  The last square brackets is  the  
``molecular interaction"  derived  in \cite{SSV_mix}.  There
are no two-fermion terms, because the integral  over the orientation of the
molecule in color space is zero. 

   One can view this Lagrangian is a variant of
the (nonlocal) Nambu-Jona-Lasinio model  at finite temperature, with 
particular coefficients and with a natural cutoff given  by the nonlocality of
the vertices, at a scale $1/\rho$. We  can now treat this interaction
analogously to refs. \cite{NJL1,NJL2} for the  NJL model. 

\section{The Mean Field Approximation and Thermodynamics  }
 To calculate the thermodynamic properties of the system, we 
first bosonize the fermionic action, making
Hubbard-Stratonovitch transformation,
\be
\sqrt{\prod \alpha(k_i)}(\psi^\dagger \psi)^2 \to 2 
\sqrt{\alpha(k_1)\alpha(k_2)}\psi^\dagger(k_1) 
\psi(k_2)V_4(2\pi)^4\delta^4(k_3-k_4)Q- \cr  V^2_4(2\pi)^8
\delta^4(k_1-k_2)\delta^4(k_3-k_4)Q^2.
\ee
Then we consider the bosonic 
field $Q$  as a constant, to be determined from free energy 
minimization.
As there is no net topological charge,  $\beta_+=\beta_-=\beta,$. We
also define 
$f={2N_m \over N_{tot}}, N_+=N_-={1-f\over 2}n_{tot}V_4, 
N_m={f \over 2}n_{tot}V_4.$
After integration over the fermion degrees of freedom we 
have the following free energy\footnote{
      Note that, our free energy is not complete in the following sense:
the gauge fields are assumed to be only a superposition of instantons, while
all excitations of the gluonic degrees of freedom are  excluded. 
In principle,  there should be a gluonic term (similar to
the the last quark term in the action above), which eventually (at high T)
will lead to the perturbative gluonic part of the thermal energy. We do not 
include it because  (i) Glueballs
 are much heavier 
than mesons and constituent quarks, and are not excited at $T\sim T_c=150 MeV$.
(ii) We mainly use the free energy in order to 
determine parameters 
such as  $f$ and the quark- 
related quantities $Q, \beta, \beta_m$.  The missing term 
describing gluonic excitations can hardly depend  on them.
 }:
\be
{\Delta F \over V_4}= -T^2_{I\bar I} \beta_m  
+(1-f)n_{tot} \log ({\beta \over c_{\rho}})+{f \over 2}n_{tot} \log 
({\beta_m  \over c_{\rho}^2 c}) + 
{Q^2 \over 8 N_c^2}(\beta+2 \beta_m) - \cr 
4N_c\int {d^4k\over(2 \pi)^4}\log(k^2+M^2(k)),  
\ee
where one can identify the chemical potentials for the ``liquid" 
component $\log ({\beta \over c_{\rho}})$ and for the ``molecular" one 
$\log({\beta_m \over c_{\rho}} c),$.
The last term corresponding  to a gas\footnote{We remind that
the instanton vacuum has no confinement, so quarks just
change their mass in the transition.} of massive quarks with the 
(instanton-induced) momentum-dependent mass
\be
M(k)={\alpha(k) Q(\beta+2\beta_m) \over 4N_c^2}
\ee
We shall determine this mass from a self-consistency condition, represented 
graphically in Fig. 2 (a), where the vertex {\cal K} is a sum of the instanton 
and the molecular interactions (Fig. 2 (b)).

   Our next step is minimization of  $\Delta F$ 
with respect to $f,\beta_m, \beta,Q.$ There are four equations:
\be
\beta_m- c \beta^2  &=&0,\\  
T^2_{I \bar I}  -{fn_{tot} \over 2\beta_m}+8N_c\int 
{d^4k\over(2 \pi)^4}{2 M(k)\over k^2+M^2(k)}{Q \alpha(k) \over 4N_c^2} - 
{Q^2 \over 4 N_c^2} &=&0,\\ 
-{(1-f)n_{tot} \over \beta}+4N_c\int 
{d^4k\over(2 \pi)^4}{2 M(k)\over k^2+M^2(k)}{Q \alpha(k) \over 4N_c^2} - 
{Q^2 \over 8 N_c^2} &=& 0, \\ 
4N_c\int {d^4k\over(2 \pi)^4}{2 M(k)\over k^2+M^2(k)}{\alpha(k) 
\over 4N_c^2} - {2Q \over 8 N_c^2}&=&0.
\ee

 One can notice, that the first equation is the condition for 
chemical equilibrium between the random and molecular components. 
These four equations can be reduced to one ``gap" equation, 
that has to be solved 
numerically: 
\be
{1\over N_c}\int {d^4k\over(2 \pi)^4}{(1+2 c \beta) 
\alpha^2(k) \over k^2+(1+2 c \beta)^2 (1-f) n_{tot}\alpha^2(k)/(2 N_c^2)}=1,
\ee
with:
\be
\beta=\big[-(1-f)n_{tot} +\sqrt{(1-f)^2n_{tot}^2  
+f/(2c)n_{tot} T_{I\bar I}^2}\big]{1 \over T_{I\bar I}^2}.
\ee

  At $T_c$, all	unpaired instantons disappear $(f \to 1)$ and the chiral 
symmetry is restored. The condition $f=1$ is satisfied on a line in the 
 $T,S_0$ plane . We have chosen $T_c=150 MeV =0.26/\rho$ and
$S_0=9.5$.
Of course, when $f=1$ or $f=0,$ the saddle point approximation for the 
$\beta$ integrals is no longer valid, so our model is restricted in the 
region $0<f<1,$ where the saddle point method is justified in the 
thermodynamic limit $V_4\to \infty.$
We see that when $f\to 1 $, $\beta_m=1/2 n_{tot}/(iT_{I\bar I})^2,$ 
$\beta=\sqrt{n_{tot}/(2 c (iT_{I\bar I})^2)} $ does not tend to 0, so that 
the $U_A(1)$ symmetry remains broken when $T \to  T_c $. This can be seen also 
in our discussion of the correlation functions in section 4.  

Our results are presented in Fig.3, where the molecule 
fraction $f$, the effective quark mass $M({\bf 0},\pi T)$, 
the quark condensate 
$\langle \bar \psi \psi \rangle$ and the 
energy density $\epsilon$  are presented. Note, that the local 
 $\langle \bar \psi \psi \rangle$ has one power of the formfactor 
$\alpha$  less under the momentum integral, compared to the nonlocal mean 
field $Q$. $\langle \bar \psi \psi \rangle$ is normalized to the 
phenomenological value of the quark condensate at $T=0$. 
We see that in about $20 MeV$ the fraction of the molecules drops to about
half, the quark condensate almost reaches its phenomenological  value at $T=0$,
but the constituent quark mass, which is defined as the value of the momentum
dependant mass from the gap equation at zero spatial momentum and Matsubara 
frequency $\pi T$, is still low -- 
about 1/3 of its phenomenological  value at $T=0$. 
(The reader should be warned that one can hardly compare the mass
value obtained in the present calculation with the usual constituent quark
mass $M(\vec k=0,\omega=0)\sim 350 MeV$ because the minimal energy possible
 $\omega\approx \pi T_c$ is rather large.) In addition, the finite 
temperature formfactors  have a maximum at spatial momentum ${\bf k}_{max} \not
= 0$, so we show with a dotted line the temperature dependence of the mass
defined at $k=({\bf k}_{max} , \pi T)$.

This behaviour of the thermodynamic quantities looks similar to the one 
obtained in \cite{NVZ-ft} or  in the NJL models \cite{NJL2}, and both
are governed by similar gap equations for the constituent quark mass 
or the quark condensate. However the physics of the transition in our 
case is quite different. In \cite{NJL2} the transition is 
governed by the suppression of the quark condensate by thermal 
excitations of quarks. In 
\cite{NVZ-ft} there is an additional thermal suppression of the 
instanton density, which results in additional suppression of the effective
coupling constant. In our case, the main effect is the 
reorganization of the instanton vacuum: it leads to reorganization of the
Lagrangian by itself. This can be seen by the fact 
that the mean field $Q $ (a $nonlocal$ analogue of the quark condensate), 
is proportional to square root of the density of the random component $1-f$. 
Our mechanism makes the transition region narrower and the transition 
temperature $T_c$ lower.

In Fig. 4 (a) we show the pressure $p=-\Delta
F$ (the solid line), together with its different components: 
pressure of free massless fermions $4N_c\int 
{d^4k\over(2 \pi)^4} \log(k^2)$ (open triangles), the deviation from
it, due to the effective mass $4N_c\int {d^4k\over(2
\pi)^4}\log({k^2+M^2(k)\over k^2})$
(open squares), the contribution from the condensate $-{Q^2 \over 8 N_c^2}
(\beta+2 \beta_m)$ 
(stars), the contributions from the molecules $T^2_{I\bar I} \beta_m 
-{f \over 2}n_{tot} \log ({\beta_m  \over c_{\rho}^2 c})$ (black squares),
and the instanton liquid $-(1-f)n_{tot} \log ({\beta \over c_{\rho}})$ 
(black triangles). 
For comparison,we also show that the pressure of a pionic gas $p=0.987 T^4$
(the dotted line) is significantly
smaller than any of the components under consideration, and thus unimportant. 

In Fig. 4.(b) the energy density $\epsilon$ below the phase 
transition point is shown to be actually directly proportional to the  molecule 
fraction $f$. This correlates well with the observation made 
in \cite{SSV_mix}, that (unlike the individual instantons) the molecules 
have a net positive energy, even in the classical approximation.

\section{ Mesonic Correlation Functions }

  Our last step is investigation of the effect of the 4-fermion
  effective interaction on mesonic spectra at $T\approx T_c$. Using
 Bethe-Salpeter equation one may calculate mesonic 
corelation functions, similarly to what was done in \cite{NJL1,NJL2}. 
Our major advantage  is that we naturally have a non-local 
vertices, which provide an ultraviolet cutoff. We start with the two-body 
interaction kernel {\cal K}, which is given by the four-fermion terms in 
the effective lagrangian. Then we have the BS equation for the 
quark-antiquark T matrix(Fig. 2 c) (See also Fig. 2 (c)):
\be
{\cal T}(q)= {\cal K}+i {\rm Tr} \int {d^4P \over(2 \pi)^4} \alpha(p+{q\over 2})
\alpha(p-{q\over 2})[{\cal K}S_F(p+{q\over 2}){\cal T}(q)S_F(p-{q\over 2})].
\ee
The trace is taken over the Dirac, flavour and color matrices. For the 
colorless meson channels we need only the color singlet terms in the 
lagrangian. Then  using the symmetries of the Matsubara space-time, we can 
decompose {\cal T} and {\cal K} into covariant structures.
 
\be
\langle \bar q_4 q_3 | {\cal K}|\bar q_2 q_1 \rangle= 
\sum_{i,\alpha}K^i_\alpha \left[ \Gamma_\alpha {\tau^i \over 
2}\right]_{34}\left[ \Gamma_\alpha {\tau^i \over 2}\right]_{12},
\ee
where $\Gamma \alpha$ denotes Dirac tensors.

Because at $T \not = 0$ there is less symmetry than at zero temperature, we 
have more structures. However our particular action (15) contains only 
scalar and pseudoscalar terms and the time oriented vector and axial terms, 
produced by $\gamma_0$ and$\gamma_0 \gamma_5$, therefore only the 
following coefficients are non-zero:
\be
K_S^i&=&{-1 \over 4N_c^2}[2 \beta_m+2(\delta_{i0}-{1\over 2})\beta], \cr
K_P^i&=&{1 \over 4N_c^2}[-2 \beta_m+2(\delta_{i0}-{1\over 2})\beta], \cr
K_V^i&=&{1 \over 2N_c^2}\beta_m, \cr
K_A^i&=&{1 \over 2N_c^2}(1-4 \delta_{i0})\beta_m
\ee
If we define the loop integral:
\be
J^{ij}_{\alpha  \beta}(q)= iN_c \tr \int {d^4p \over(2 \pi)^4} 
\alpha(p+{q\over 2})\alpha(p-{q\over 2})[\Gamma_\alpha 
{\tau^i \over 2} S_F(p+{q \over 2}) \Gamma_
\beta {\tau^j \over 2} S_F(p-{q \over 2})],
\ee
then the solution of the BS equation (in matrix notation) is:
\be
T=[1-JT]^{-1}K.
\ee
To get the mesonic correlation function for the corresponding channel, 
we have to multiply T from the left and the right with two loop  
integrals $J$ (those have only two formfactors $\sqrt{\alpha}$, instead 
of four, because the correlation function is defined for point-like 
currents). For all channels, we have separated equations except 
for the pseudoscalar and axial vector channels which mix 
both for isospin 1 ($\pi$) and 0 ($\eta$).  In Fig.5 we show the 
Fourier transforms of the 
correlation functions for Euclidean time separation of the currents,
normalized to the free correlation function at finite temperature. We 
show them in 6 steps with $\Delta T \approx 9 MeV$, starting from $T = 
150 MeV$ and ending 
with $T \approx 105 MeV$. 

The most striking feature is the strong 
attraction in the pion channel, which remains robust and does not disappear 
at $T_c$. This is due mainly to the residual 't Hooft interaction 
(induced by the propagating quarks) and in a lesser extent to the attractive 
character of the molecular interaction in this channel (triangles). This 
feature is in agreement with the lattice simulations around 
$T_c$ \cite{Karsch'94}, which 
also show a strong pion signal beyond $T_c$. Futhermore, at $T_c$ our pion 
signal is similar to the one from the numerical simulations of the interacting 
instanton liquid \cite{SS_95b}, although stronger\footnote{ The differences 
might be due to the fact that we are considering the chiral limit, while in 
\cite{SS_95b} the quarks have a nonzero mass, which leads to the smearing or 
all signals.}. 

The results are in agreement with all chiral
theorems, and they clearly show restoration of the chiral symmetry at $T_c$.
In particular, at $T=T_c$ the $\pi$ pseudoscalar correlation 
function $\langle PP^1 \rangle$ coincides with the $\sigma$ scalar correlation 
function $\langle SS^0 \rangle$.  Another feature 
in agreement with the chiral symmetry is that the pion is 
decoupled from the axial current $(f_{\pi} \to 0)$ at $T \to T_c$,  but 
remains coupled to the pseudoscalar one $(\lambda_{\pi} \not = 0)$ even in 
the pure molecular vacuum (f=1). 

  An open theoretical issue debated in the literature is how strongly the 
 $U_A(1)$ chiral symmetry
is violated at $T \sim T_c$, see e.g. \cite{Shuryak_com,Cohen,Hatsuda,Hsu}.
We remind that this symmetry is strongly broken by  
random liquid at $T=0$, but it is respected by the new term in the Lagrangian 
due to ``molecules" which we derived above. Although
molecules are prevailing above $T_c$ the 
't Hooft interaction 
does not disappear completely  at any T.
One way to explain  it, is to say that 
 external quarks induce additional instantons, absent in vacuum. 

 The  way to measure  $U_A(1)$-violating effects is to 
calculate the 
correlation function for $\eta$ (the isoscalar pseudoscalar) or the isovector 
scalar, to be referred to as $\delta$,
and compare their  properties with the pion or sigma ones.
These two channels are  
$U_A(1)$ partners of $\sigma$ and $\pi$, and if this symmetry gets
approximately restored, those should converge.

 As shown in 
\cite{SSV}, in the random instanton vacuum the correlation functions in 
the $\eta,\delta$  channels, display so strong repulsive interaction 
that they become negative\footnote{This is an artefact of the 
``randomness" of the instanton liquid  related to too 
strong fluctuations of the topological charge: it disappears in the
interacting
instanton liquid.}. Our results show that although
the $\eta, \delta$ correlators {\it increase} when T 
approaches the phase transition point, the $U(1)$ restoration still does not 
happen in our model. For comparison, we have plotted   in Fig. 5 (triangles)
the same correlation functions at $T_c$ if  only  the $U(1)$ symmetric
molecule-induced interaction is included.

Because we obtain the correlation functions in for 
Euclidean momenta by numerical procedure, we cannot 
analytically continue them and find the pole
meson masses. However, the $\pi$ mass is guaranteed by the Goldstone theorem 
to be 0. We can therefore derive  the pion decay constants $f_\pi$ and 
$\lambda_\pi$. The quark T matrix in the pseudoscalar, $I=1$ channel
has the following form near the pion pole (see Fig. 2 (d)):
\be
 T_\pi |_{q_0=0,{\bf q} \to 0}=
{[g_P(i \gamma_5 \tau_3)+g^0_A(-i \gamma_0 
\gamma_5 \tau_3)] \tensor [g_P(i \gamma_5 \tau_3)+
g^0_A(i \gamma_0 \gamma_5 \tau_3)] \over q_0^2+{\bf q}^2}
\ee
Comparing with (30), we get:
\be
g_P&=&\sqrt{K_P^3(1-K_A^3J^{33}_{A^0 A^0}) \over dD_\pi/dq^2}
\Big|_{q_0=0,{\bf q} 
\to 0}, \\ 
g_A&=&g_P {K_A^3 J^{33}_{P A^0}) \over (1-K_A^3J^{33}_{A^0 A^0})}
\Big|_{q_0=0,{\bf q} \to 0},
\ee
where $D_\pi=\det(1-K_\pi J_\pi)=(1-K_P^3 J_{PP}^{33})(1-K_A^3 
J_{AA}^{33})-K_P^3K_A^3(J^{33}_{PA})^2$.
Using the definitions:
\be
i f_\pi q^\mu \delta_{ij}&=& \langle 0| \bar \psi(0) \gamma^\mu \gamma_5 
{\tau^i \over 2} \psi(0)| \pi_j(q) \rangle,  \\ 
\lambda_\pi \delta_{ij}&=& \langle 0| \bar \psi(0) i \gamma_5 
{\tau^i \over 2} \psi(0)| \pi_j(q) \rangle,
\ee
and calculating a simple loop diagram (Fig 2. (d)), we get:
\be
\lambda_\pi &=& \tilde T_{PP}^3 g_P,  \\ 
f_\pi^0 &=& (\tilde T_{PA^0}^3/q^0 g_P+\tilde T_{A^0A^0}^3 g_A/q^0)
|_{q_0=0,{\bf q} \to 0}, \\ 
f_\pi^i &=& (\tilde T_{PA^i}^3/q^i g_P)|_{q_0=0,{\bf q} \to 0},
\ee
where the tilde indicates that the loop integral is with two factors of 
$\sqrt{\alpha}$ only, and because of the lack of $O(4)$ symmetry, there 
are two $f_\pi$'s, coupled to the time and the spatial components of the 
axial current. The results are shown in Fig. 6. $g_A$ and the two $f_\pi$'s 
go to 0 at $T_c$ as required by the restoration of the chiral symmetry. 
$\lambda_\pi$ however, remains finite.  This means that the pion
(and also his chiral partner, $\sigma$), survive the phase transition! 
This conclusion is consistent with numerical evidences obtained from the
calculation of the Euclidean correlation functions in the time direction 
\cite{SS_95b}, but there $\lambda_\pi$ although remaining  finite, decreases 
towards $T_c$,  while in our calculation 
$\lambda_\pi$ slightly increases. This difference, as we have mentioned 
before, might be due to the non-zero quark current masses in \cite{SS_95b}.

  Furthermore, in our approach we have found a signal of attractive
  interaction also in the vector ($\rho$) channel. It is seen as an additional
  maximum in the correlators, shown in Fig.5 (i), which is the same as the one,
generated by the molecule-induced
interaction (triangles). This signal was $not$ observed in
\cite{SS_95}\footnote{One possible explanation is related to the different
treatment of the non-zero mode part of the quark propagator. 
In  \cite{SS_95} an additional repulsive interaction is included. 
The same authors have done simulations with the free propagator, that we use, 
for the non-zero mode of the quark propagator and they have noticed an 
attractive correlation function in the $\rho$ channel. However 
some other contributions are not 
included in both approaches (e.g. the confinement, which provides an 
attractive interaction).
So the question, whether $\rho$ ``melts" below or at $T_c$, remains open.}.

\section{Summary} 
   In this  paper we have  studied individual
   instanton-antiinstanton molecules,  
in much greater  details than it was done before. We have determined 
classical bosonic interaction in the ratio ansatz and quark-induced
interaction. Then we have performed 11-dimensional integration over
collective variables, 10 in the saddle-point approximation and the
last one explicitly.

  The results obtained, have been
used in studies of molecule formation in the ensemble, at temperatures
close to chiral restoration point $T\approx T_c$. We have used
a two-component (or ``cocktail") model, with contributions from uncorrelated 
instanton liquid and polarized $I\bar I$ molecules,
 and have confirmed that chiral restoration is driven by
formation of the instanton-antiinstanton molecules.
 
 Both random and molecular components have generated an effective 4-fermion 
interaction. With
standard mean-field methods we have derived  semi-analytically
the thermodynamics of the system. The
basic conclusion is that there is a rapid temperature dependence
 of the fraction of molecules f, see Fig.3(a),
which jumps from $f\sim 0.5$ at $T=0.7 T_c$ to 1 at $T_c$. The corresponding 
jump in energy density is strongly correlated with it (see Fig.4(b)), confirming
the idea suggested in \cite{SSV_mix} that
formation of ``molecules" is the major reason of 
 rapid growth of the energy density around the phase transition point.  
                                                  
We have also calculated the mesonic correlation functions,
using Bethe-Salpeter type equation. We have found that the pion is 
decoupled from the axial current at $T_c$, as it should, 
but remains coupled to 
pseudoscalar one even in pure molecular vacuum (f=1). A strong signal,
similar to that reported numerically in \cite{SS_95b}, has been found,
indicating that pion may survive the phase transition as a bound state.
 Sigma meson  follows 
the pattern of chiral symmetry restoration, joining the pion. The 
``repulsive" channels $\eta,\delta$ show increase of the correlators,
but they do not exactly join the pion and the sigma ones, so $U(1)_A$ 
symmetry remains broken.

\section{Acknowledgements}
  We would like to thank J.~J.~M.~Verbaarschot for the many useful
discussions. The
reported work was partially supported by the US DOE grant
DE-FG-88ER40388.

\newpage\noindent
{\Large\bf figure captions}\\ \\ \\
\underline{figure 1}  (a) The time dependence of $Z_{mol}'=\int 
d^{10}\Omega 
T^4_{I \bar I}e^{-S_{int}}$, (the integration over the Matsubara time is
missed) at $T_c$,
(b) the time dependence of $T_{I \bar I}$, at $T_c$,
(c) the temperature dependence of $Z_{mol}$, (d) the temperature dependence 
of $T_{I\bar I}$. 
\\ \\
\underline{figure 2}  (a) The Hartree-Fock (self-consistency) equation for 
the quark propagator. 
(b) the four fermion interaction is a sum of the 't Hooft vertex and the 
molecular vertex.
(c) The Bethe-Salpeter equation for the {\cal T} matrix in the mesonic 
channels.
(d) The pion {\cal T} matrix near the massless pion pole.
(e) The diagram for the pion coupling to the axial current.
\\ \\
\underline{figure 3}  (a) The molecule 
fraction $f$, (b) the effective quark mass $M({\bf 0},\pi T)$ (solid line) and 
$M({\bf k}_{max},\pi T)$ (dotted line), (c) the 
energy density $\epsilon$ and (d) the quark condensate 
$\langle \bar \psi \psi \rangle$, normalized to the phenomenological 
value at zero temperature  as a function of the temperature. All 
dimensional quantities are in units of the inverse instanton radius 
$1/\rho = 580MeV$.
\\ \\
\underline{figure 4} (a) The pressure $p=-\Delta F$,
 and the different components of it, as a function of the temperature. 

The pressure is shown by a solid line, and the different contributions to it 
are as follows: pressure of free massless fermions $4N_c\int 
{d^4k\over(2 \pi)^4} \log(k^2)$ (open triangles), the deviation from
it, due to the effective mass $4N_c\int {d^4k\over(2
\pi)^4}\log({k^2+M^2(k)\over k^2})$
(open squares), the contribution from the condensate $-{Q^2 \over 8 N_c^2}
(\beta+2 \beta_m)$ 
(stars), the contributions from the molecules $T^2_{I\bar I} \beta_m 
-{f \over 2}n_{tot} \log ({\beta_m  \over c_{\rho}^2 c})$ (black squares),
and the instanton liquid $-(1-f)n_{tot} \log ({\beta \over c_{\rho}})$ 
(black triangles). For comparison, the pressure of a pion gas is shown by
a dotted line.

The the dependence of the energy density $\epsilon$ on the molecule 
fraction $f$ (fig. 4b).
\\ \\

\underline{figure 5} Mesonic 
correlation functions for Euclidean time separation of the currents,
normalized to the free correlation function at finite temperature. At 
$\tau=0$ and $\tau=1/T$ they are 1, but we have subtracted 1 for convenience. 
$\tau$ is in units of $\rho$.
These are the isospin 1 pseudoscalar-pseudoscalar (a), pseudoscalar-axial 
(b) and axial-axial (c) correlators,the isospin 0 pseudoscalar-pseudoscalar 
(d), pseudoscalar-axial 
(e) and axial-axial (f) correlators, the scalar-scalar correlators for 
isospin 1 (g) and 0 (h), and the vector-vector correlators,which is the 
same for isospin 1 and 0 (i). There 
are 6 different temperatures, starting from $T \approx 150 MeV$ and ending 
with $T \approx 105 MeV$.  For comparison, the correlation functions of a 
purely ``molecular" vacuum at $T_c$ are plotted with triangles. 
\\ \\
\underline{figure 6} The effective pion-quark couplings $g_P$ and 
$g_A$ (a) and the pion constants $\lambda_\pi$ , $f_\pi^0$ and
$f_\pi^i$ (b) as a function of the temperature. All 
dimensional quantities are in units of the inverse instanton radius 
$1/\rho = 580MeV$.

\newpage
\begin{figure}
\begin{center}
\leavevmode
\epsffile{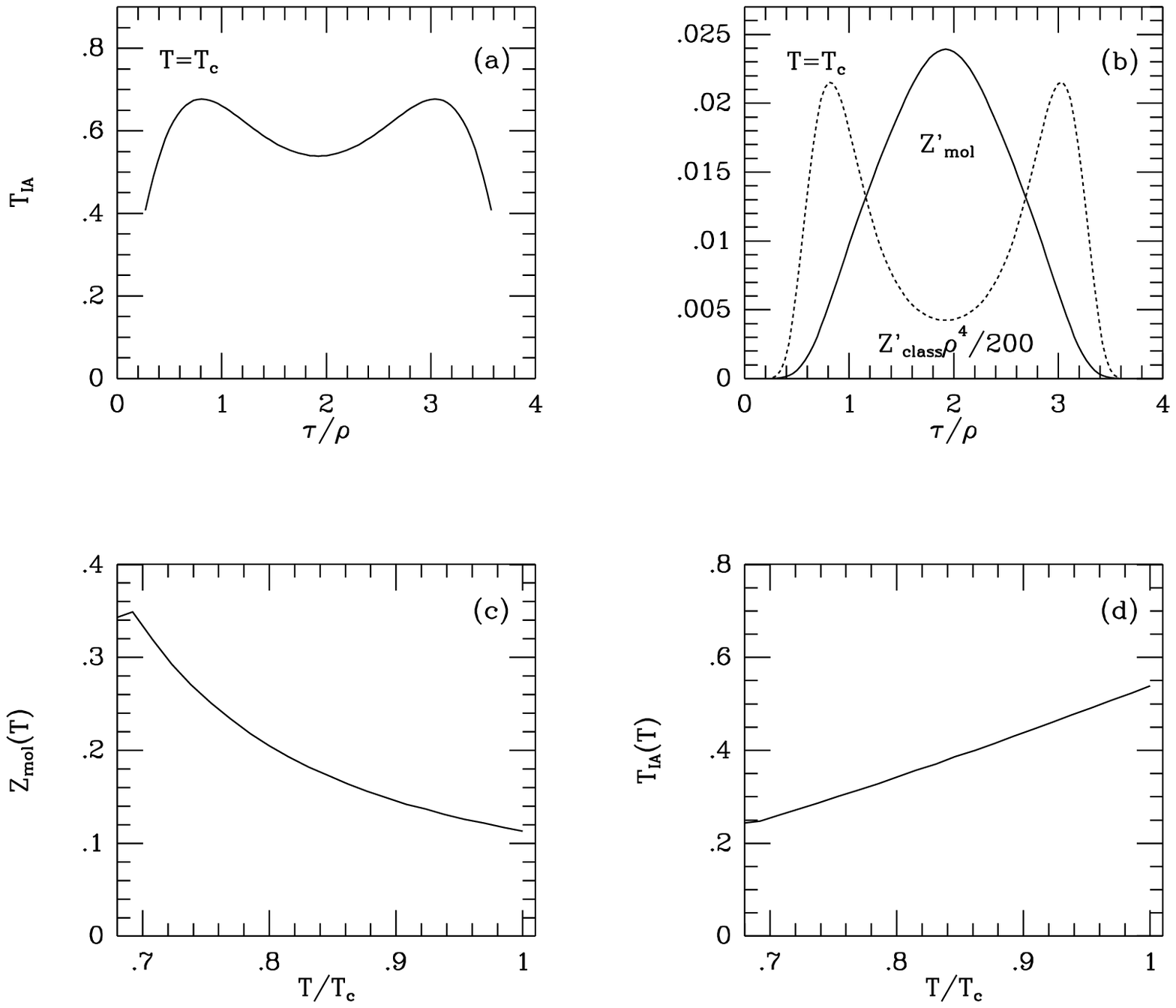}
\end{center}
\caption{}
\end{figure}
\vfill

\newpage
\begin{figure}
\begin{center}
\leavevmode
\epsffile{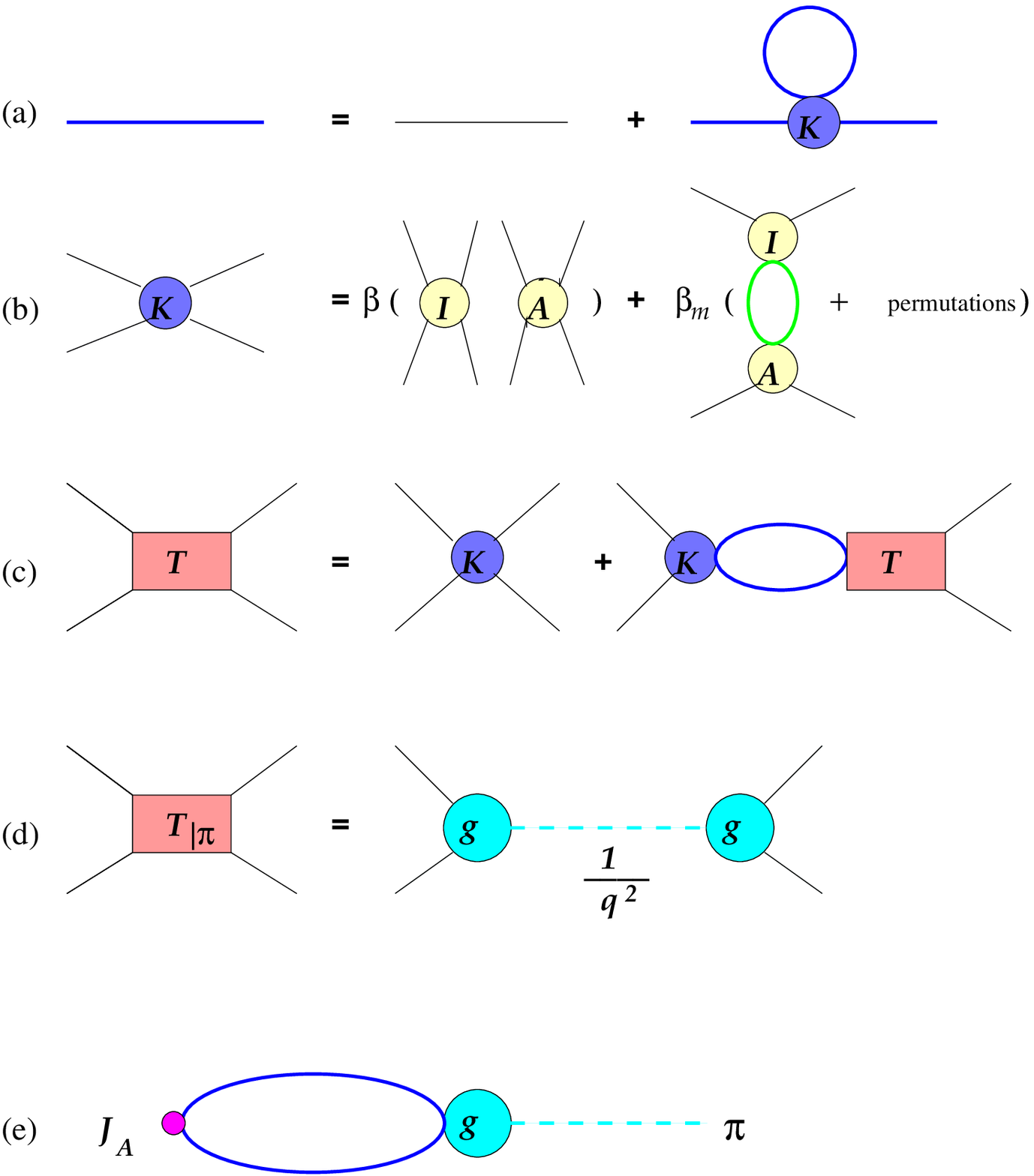}
\end{center}
\caption{}
\end{figure}
\vfill

\newpage
\begin{figure}
\begin{center}
\leavevmode
\epsffile{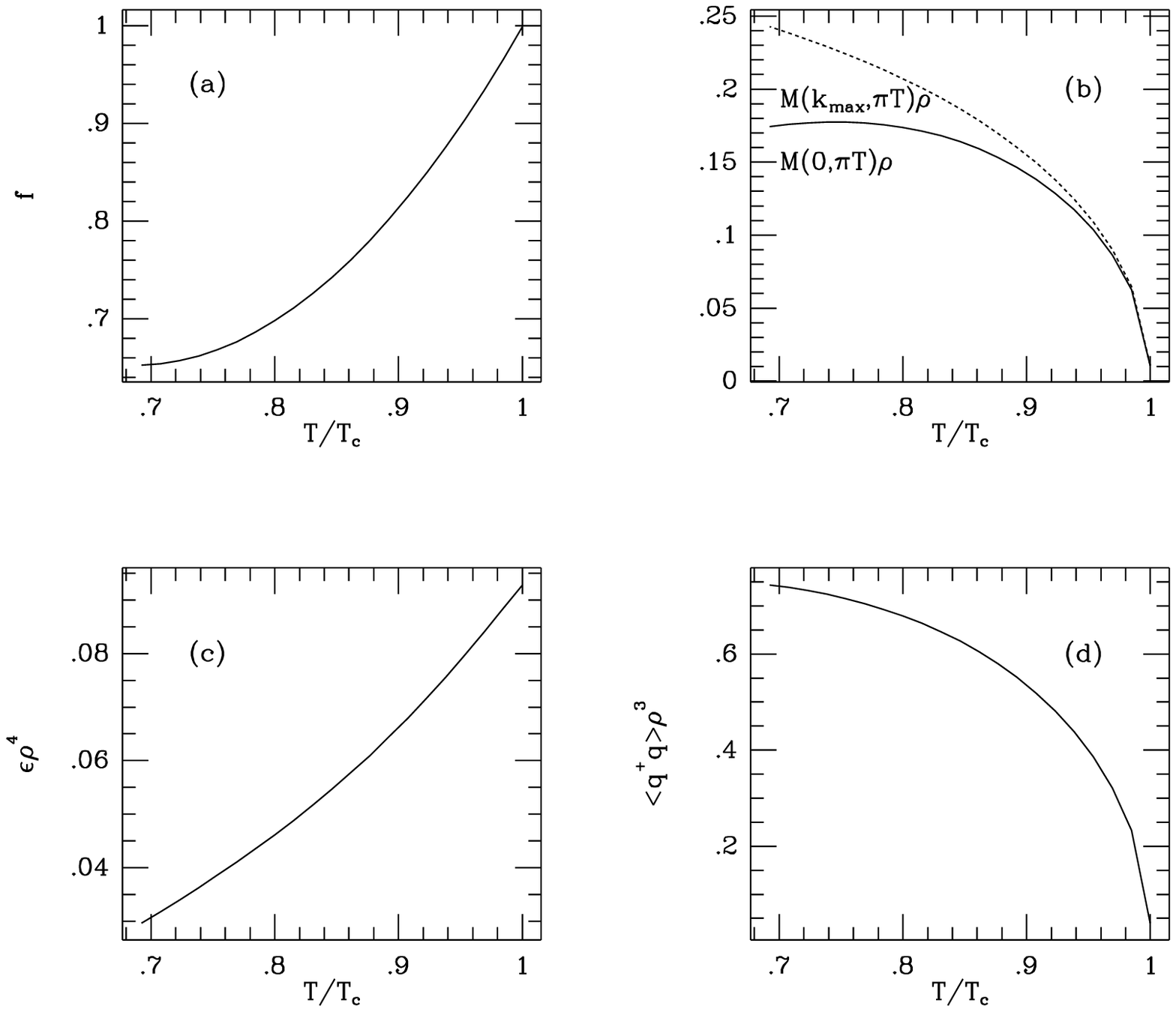}
\end{center}
\caption{}
\end{figure}
\vfill

\newpage
\begin{figure}
\begin{center}
\leavevmode
\epsffile{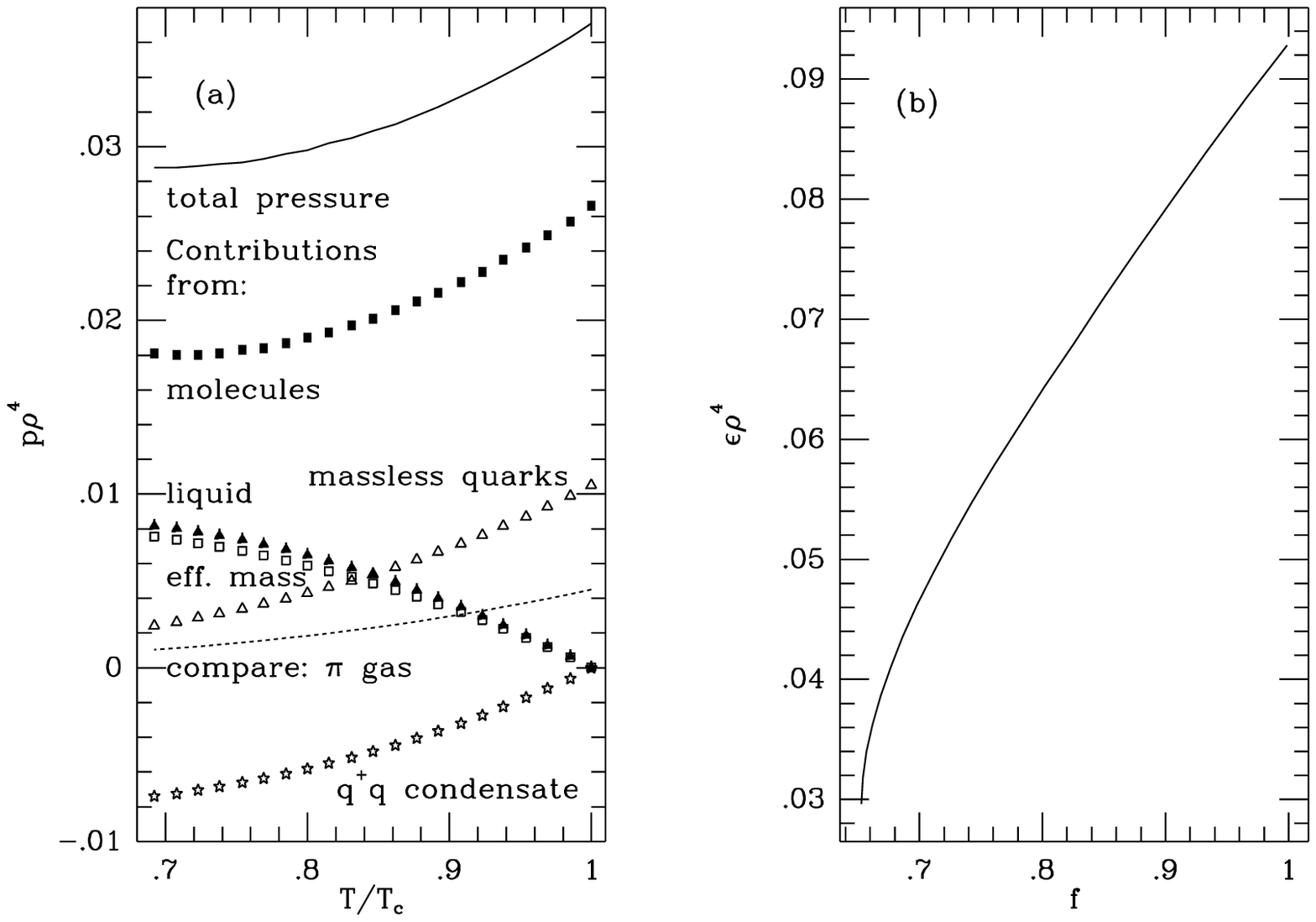}
\end{center}
\caption{}
\end{figure}
\vfill

\newpage
\begin{figure}
\begin{center}
\leavevmode
\epsffile{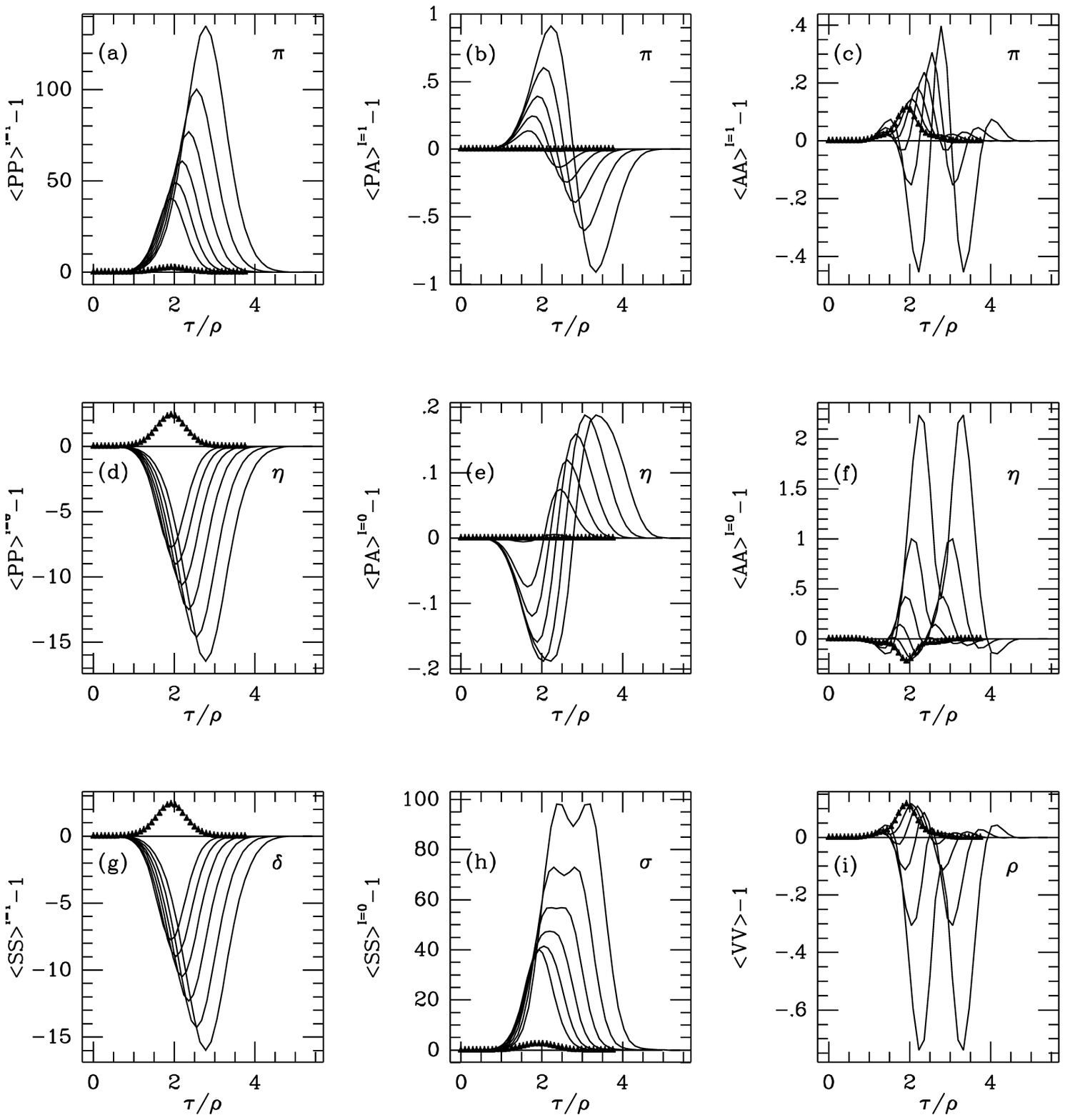}
\end{center}
\caption{}
\end{figure}
\vfill

\newpage
\begin{figure}
\begin{center}
\leavevmode
\epsffile{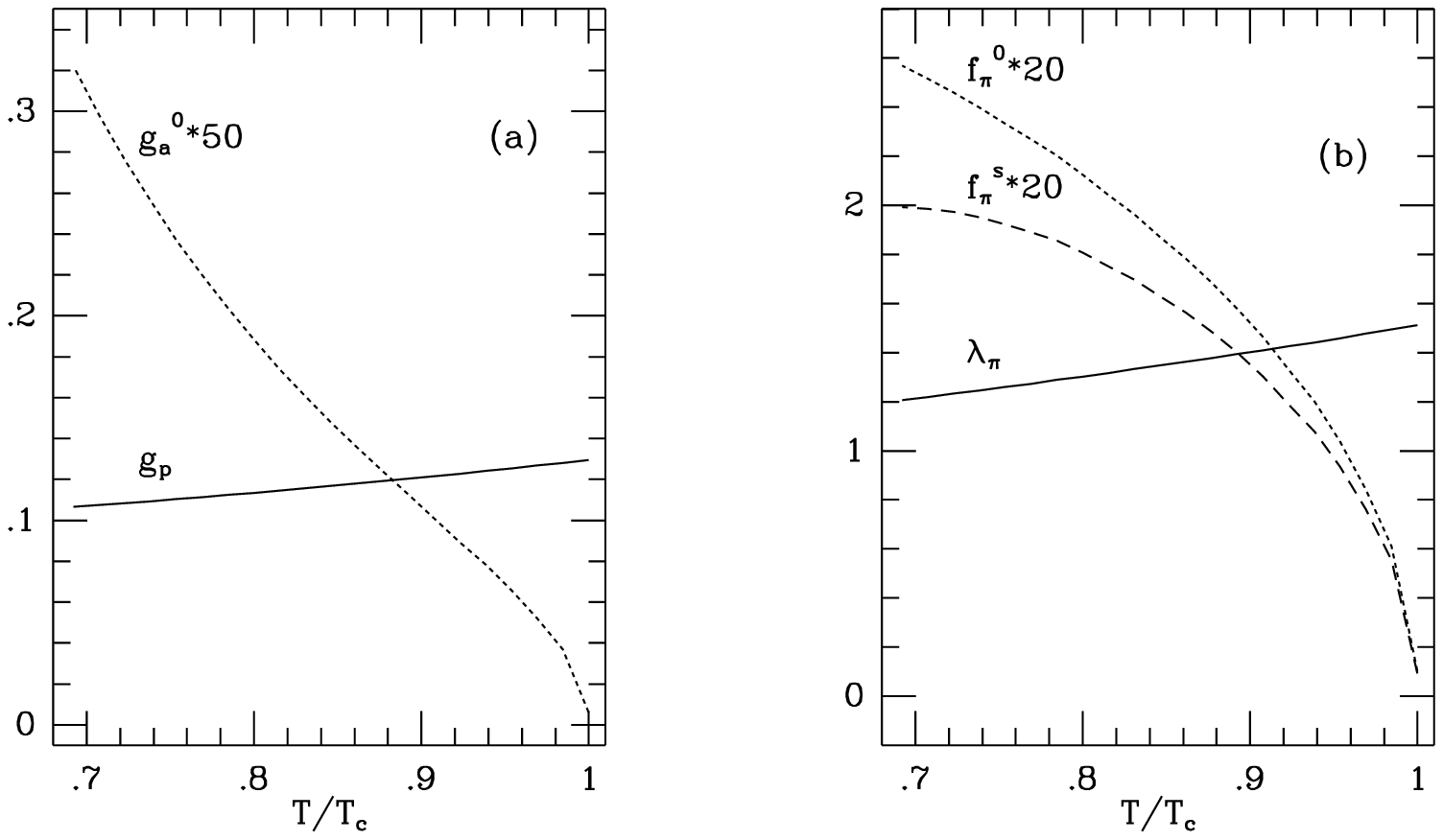}
\end{center}
\caption{}
\end{figure}
\vfill


\newpage
\begin{thebibliography}{xx}

\bibitem{Karsch}
F.~Karsch, plenary talk at Lattice-93,{\it Nucl.~Phys.~(Proc.~Suppl.)} 
{\bf B34} (1994) 63.
\bibitem{Detar}
C.~DeTar,  plenary talk at Lattice-94,{\it Nucl.~Phys.~(Proc.~Suppl.)}{\bf B}. 
In press. Quark Gluon Plasma in numerical simulations of lattice QCD,
to appear in ''Quark Gluon Plasma 2", R.~Hwa, ed.,
World Scientific (1995)
 
\bibitem{Shu_82}
E.~V.~Shuryak,
{\it Nucl.~Phys.} {\bf B203}, 93 (1982) 116. 
 
\bibitem{DP}
   D.~I.~Diakonov, V.~Yu.~Petrov {\it Sov.Phys.Jetp} {\bf 62} 
( 1985) 204-214. ({\it Zh.~Eksp.~Teor.~Fiz.} {\bf 89} ( 1985)
   361-379).
    {\it Nucl.~Phys.} {\bf B272} (1986) 457.


   
\bibitem{NVZ}
 M.~A.~Nowak, J.~J.~M.~Verbaarschot, I.~Zahed {\it Nucl.~Phys.} {\bf B324} 
(1989) 1.


\bibitem{Shu_88}
E.~Shuryak,
{\it Nucl.~Phys.} {\bf B302} (1988) 559, 574, 599,
{\it Nucl.~Phys.} {\bf B319} (1989) 521, 541.

\bibitem{SV_90} 
E.~V.~Shuryak, J.~J.~M.~Verbaarschot {\it Nucl.~Phys.} {\bf B341} (1990) 
1-26.

   
\bibitem{SSV}
 E.~V.~Shuryak, J.~J.~M.~Verbaarschot {\it Nucl.~Phys.} {\bf B410} (1993) 55-89.
T.~Sch\"afer,
 E.~V.~Shuryak, J.~J.~M.~Verbaarschot {\it Nucl.~Phys.} 
~{\bf B412} (1994) 143;
T.~Sch\"afer, E.~V.~Shuryak, {\it Phys.~Rev.} {\bf D50} (1994) 478.
                

\bibitem{SS_95}
T.~Sch\"afer, E.~V.~Shuryak,
{\it Phys.~Rev.~Lett.} {\bf 75} (1995) 1707.

\bibitem{SS_95b}
T.~Sch\"afer, E.~V.~Shuryak,
{\it Phys.~Lett.} {\bf B356} (1995) 147.


\bibitem{Shuryak_cor} E.~Shuryak, {\it Rev.~Mod.~Phys.} {\bf 65} (1993) 1.


\bibitem{Negele}
M.~C.~Chu, J.~M.~Grandy, S.~Huang and J.~W.~Negele,
{\it Phys.~Rev.~Lett.} {\bf 70} (1993) 225;
M.~C.~Chu and S.~Huang, {\it Phys.~Rev.} {\bf D45} (1992) 7.

\bibitem{CGHN_94}
M.~C.~Chu, J.~M.~Grandy, S.~Huang, J.~W.~Negele,
{\it Phys.~Rev.} {\bf D49} (1994) 6039.

\bibitem{MS_95}
C.~Michael, P.~S.~Spencer,
{\it Nucl.~Phys.~(Proc.~Suppl.)} {\bf B42} (1995) 261;
{\it Phys.~Rev.} {\bf D50} (1995) 7570.

 
\bibitem{Shuryak_78} E.~Shuryak
Phys.Lett. B79 (1978) 135.
\bibitem{PY} R.~D.~Pisarski and L.~G.~Yaffe,
{\it Phys.~Lett.} {\bf B97} (1980) 110.


\bibitem{IS} E.~M.~Ilgenfritz and E.~Shuryak, {\it Nucl.~Phys.}
{\bf B319} (1989) 511.
   

\bibitem{SMV} 
E.~Shuryak, M.~Velkovsky {\it Phys.~Rev.} {\bf D50} (1994) 3323-3327.


\bibitem{CS_95}
M.~C.~Chu, S.~Schramm,
{\it Phys.~Rev.} {\bf D51} (1995) 4580.


\bibitem{IS2} 
E.~M.~Ilgenfritz, E.~V.~Shuryak {\it Phys.~Lett.} {\bf B325} (1994) 263-266.

\bibitem{SSV_mix}
T.~Sch\"afer, E.~V.~Shuryak, J.~J.~M.~Verbaarschot
{\it Phys.~Rev.} {\bf D51} (1995) 1267.


\bibitem{SV-FT1-91}   
E.~V.~Shuryak, J.~J.~M.~Verbaarschot {\it Nucl.~Phys.} {\bf B364} (1991) 
255-282.

\bibitem{SSV_95}
T.~Sch\"afer, E.~V.~Shuryak, Interacting instanton liquid
preprints, Stony Brook, SUNY-NTG-95-22,SUNY-NTG-95-23.

\bibitem{Ver_91}
J.~J.~M.~Verbaarschot
{\it Nucl.~Phys.} {\bf B362} (1991) 33.

\bibitem{DP_84}
    {\it Nucl.~Phys.} {\bf B245} (1984) 259.

\bibitem{CB} C.~Bernard {\it Phys.~Rev.} {\bf D19} (1979) 3013.


\bibitem{Shuryak_com}  Edward Shuryak {\it Comments Nucl.~Part.~Phys.} {\bf 21}
(1994) 235-248.

\bibitem{'t H} G.~'t~Hooft {\it Phys.~Rev.} {\bf D14} (1976) 3432.
   
\bibitem{NVZ-ft} M.~A.~Nowak, J.~J.~M.~Verbaarschot, I.~Zahed 
{\it Nucl.~Phys.} { \bf B325} (1989) 581.


\bibitem{GPY}
D.~J.~Gross, R.~D.~Pisarski and L.~G.~Yaffe, {\it Rev.~Mod.~Phys} {\bf 
53} (1981) 43.

\bibitem{NJL1} S.~Klimt, M.~Lutz, U.~Vogl and W.~Weise {\it Nucl.~Phys.}
{\bf A516} (1990) 429-468.


\bibitem{NJL2} U.~Vogl and W.~Weise {\it Prog.~Part.~Nucl.~Phys.}
{\bf 27} (1991) 195-272. 

\bibitem{Karsch'94}
G.~Boyd, S.~Gupta, F.~Karsch, E.~Laermann {\it Z.~Phys.} {\bf C64} (1994) 331.

\bibitem{Shuryak_com} 
E.~Shuryak {\it Comments Nucl.~Part.~Phys.} {\bf 21} (1994) 235.

\bibitem{Cohen}
T.~D.~Cohen, Univ. of Maryland preprint no. 96-060. 

\bibitem{Hatsuda}
S.~Lee and T.~Hatsuda, preprint hep-ph/9601373., Jan. 1996.

\bibitem{Hsu}
N.~Evans,S.~D.~H.~Hsu, M.~Schwetz, preprint YCTP-P3-96, Jan. 1996.

  
\end{thebibliography}
\end{document}